\documentclass[12pt,prc,aps,onecolumn]{revtex4-1}   
\usepackage{epsfig}  
\usepackage{graphicx}  
\usepackage{amssymb}
\usepackage{amsmath}
\usepackage{color}  

\usepackage{dsfont}

\begin{document}

\title{Effect of  Octupole correlations on Fission of Light Nuclei}

\author{Guillaume Scamps$^{1,2}$ and C\'edric Simenel$^{3}$}

\affiliation{$^{1}$Institut d'Astronomie et d'Astrophysique, Universit\'e Libre de Bruxelles, Campus de la Plaine CP 226, BE-1050 Brussels, Belgium \\
$^{2}$Center for Computational Sciences, University of Tsukuba, Tsukuba 305-8571, Japan \\
$^{3}$Department of Theoretical Physics and Department of Nuclear Physics, Research School of Physics and Engineering \\ Australian National University, Canberra, Australian Capital Territory 2601, Australia}

\email{gscamps@ulb.ac.be}


\begin{abstract}
Fission of $^{180}$Hg produces mass asymmetric fragments which are expected to be influenced by deformed shell-effects at N~=~56 in the heavy fragment and Z~=~34 in the light fragment [G. Scamps and C. Simenel,  arXiv:1904.01275 (2019)]. To investigate both shell-effects and to determine which one has the main influence on the asymmetry in the region of the $^{180}$Hg, we produce a systematic of Constraint-Hartree-Fock calculations in nuclei with similar N/Z ratio than the $^{178}$Pt. It is found that N~=~56 determines the asymmetry of systems in this region of the nuclear chart.
\end{abstract}


\maketitle

\begin{widetext}

\section{Introduction}

The configuration at scission is expected to strongly influence the symmetric or asymmetric nature of fission \cite{Wil74}. Indeed, the shell structure of the fragments can play an important role in stabilizing the number of neutrons and protons in each fragment. Spherical magic numbers sometimes play a role as in $^{258}$Fm \cite{Hul86}. However, due to strong Coulomb interaction, and the presence of a neck between the fragments in which strong interaction between the fragments is still present, both pre-fragments are deformed with a strong quadrupole shape or a strong octupole deformation. As a result, deformed shell effects are expected to dominate \cite{Sca18,Sca19}.

Experimental data on the fission of actinides show that the heavy fragment is centered in Z~=~54 \cite{Uni74,Sch00,Boc08} for a large number of actinides. This empirical result has been explained by the role of the octupole deformation in Ref. \cite{Sca18}. Indeed the region around the $^{144}$Ba is expected to have strong octupole deformation \cite{Lea85,Buc16,Buc17}. This octupole deformation is associated with shell gaps with Z~=~52, Z~=~56, N~=~84 and N~=~88. 

To test the universality of the effect of octupole shell structure on the asymmetry of  fission,  neutron-rich mercury isotopes have been studied with a similar approach \cite{Sca19}. The $^{180}$Hg case is particular since it was expected to fission symmetrically leading to two $^{90}$Zr which is a magic nuclei with N~=~50. However, experience reveals that $^{180}$Hg fissions asymmetrically leading to fragments around  $^{80}$Kr and $^{100}$Ru \cite{And10}. The deformed shell effects associated to this fission mode are predicted to be at N~=~56 and Z~=~34 which are the numbers associated to octupole shell effect due to the repulsion of states with $\Delta$j~=~$\Delta$l~=~3 \cite{But16}. Nevertheless, the light fragment is found to be strongly elongated. Then the Z~=~34 gap is due to a more complex deformation \cite{Sca19}.
 
  The goal of the present manuscript is to determine which of the N~=~56 and Z~=~34  shell-effects is the strongest by varying the fissioning system mass while the N/Z ratio is preserved.

\section{Constrained-Hartree-Fock calculation}

The calculations are done using a version of the \textsc{ev8} code \cite{Bon05} modified in order to have only one plane of symmetry. The self-consistent Constrained-Hartree-Fock equations augmented by the BCS pairing (CHF+BCS) are solved with a constraint on the quadrupole and octupole moment defined respectively as
\begin{eqnarray}
Q_{20} &=& \sqrt{\frac{5}{16\pi}}\int d^3r \,\rho(\mathbf{r}) (2z^2-x^2-y^2),\\
Q_{30} &=& \sqrt{\frac{7}{16\pi}}\int d^3r \,\rho(\mathbf{r}) [2z^3-3z(x^2+y^2)].
\end{eqnarray}

The interaction used is the Sly4d \cite{Kim97} Skyrme functional with a surface type of pairing with interaction strength $V_0^{nn}$~=~1256~MeV$\cdot$fm$^3$ and $V_0^{pp}$~=~1462~MeV$\cdot$fm$^3$ \cite{Sca13a}. To determine the fission valleys the following procedure has been used. First, a calculation is performed with a constraint on $Q_{20}$~=~47.3 b and different values of the $Q_{30}$ moment. Then, the octupole constraint is released and the value of $Q_{20}$ are changed gradually to explore the valley in both directions. The calculation is done until $Q_{20}$~=~0 and until the system scissions respectively when $Q_{20}$ is decreased and increased.

\section{Results}

To investigate the strength of the deformed shell effects, we determine the asymmetric fission path of the systems that have similar initial N/Z ratio to the $^{178}$Pt (1.25 $\leq N/Z \leq$ 1.3).
The value of the average number of protons and neutrons just before scission is shown in Fig. \ref{fig:res_NZ_fctA}. We see that up to A~=~186, all heavy fragments have N~$\simeq$~56, while $Z_L$ varies and show no particular stabilisation at Z~$\simeq$~34, indicating a dominance of the shell gap N~=~56 over Z~=~34.

\begin{figure}[h]
\includegraphics[width= 0.49 \linewidth]{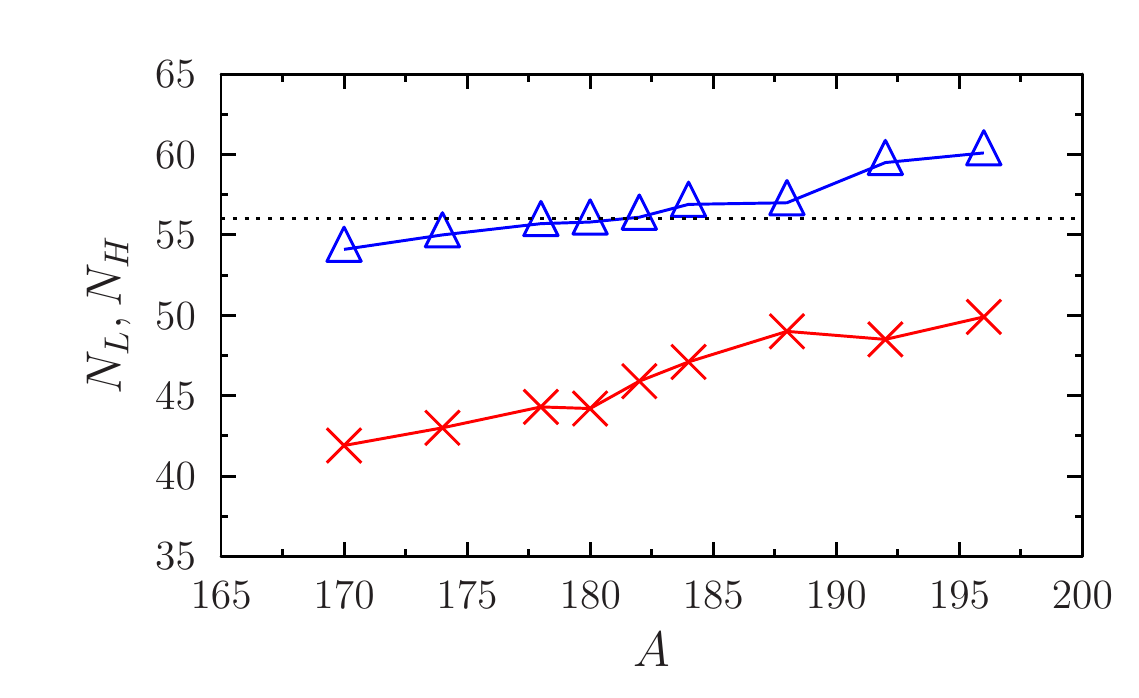}
\includegraphics[width= 0.5 \linewidth]{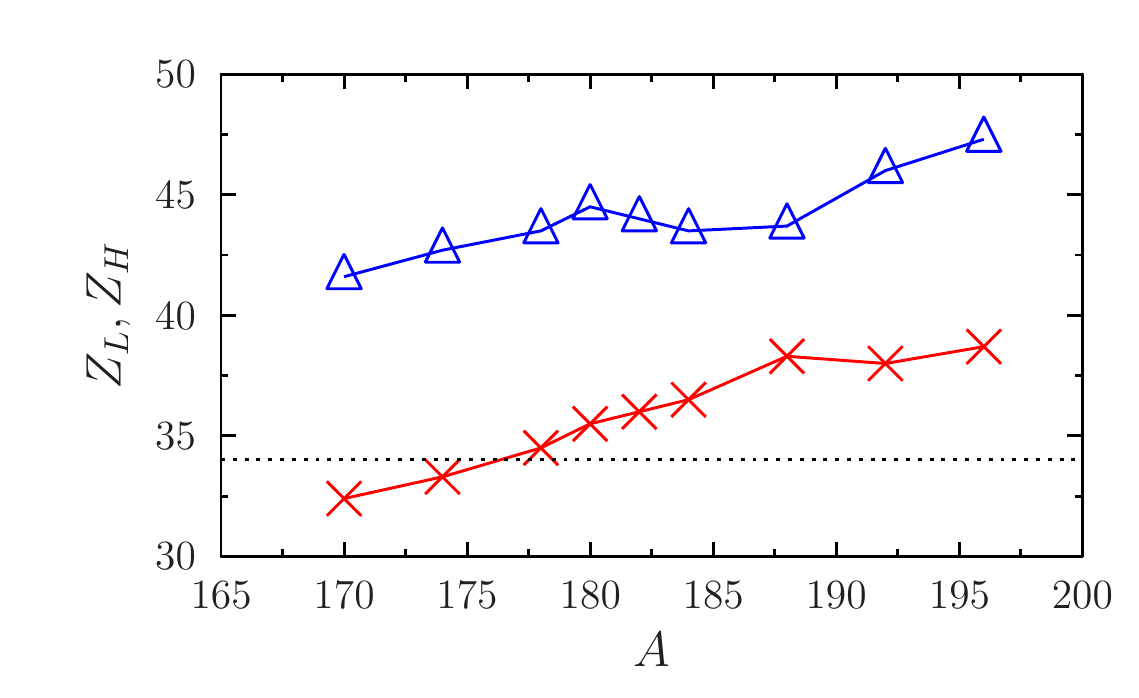}
\caption{Average neutrons (left panel) and protons (right panel) in the fission fragment as a function of the mass of the compound system. The light fragments are shown by red crosses and the heavy by blue triangles. The studied systems here are $^{170}$W, $^{174}$Os, $^{178}$Pt, $^{180}$Hg, $^{182}$Hg, $^{184}$Hg, $^{188}$Pb, $^{192}$Po and $^{196}$Rn.}
\label{fig:res_NZ_fctA}
\end{figure}

Note that in the case of the $^{188}$Pb, $^{192}$Po and $^{196}$Rn the potential energy surface as a function of asymmetry is essentially flat. This is due to the presence of different shell-gap that arise at different asymmetries. For example in $^{196}$Rn the N~=~56 shell-gap favor the symmetric fission while the Z~=~34 and the Z~=~44 favor two different asymmetric valleys. As a consequence, the shell effect does not fix strongly a particular value for the asymmetry at the scission in the 3 nuclei. 

We show in Fig. \ref{fig:PES_proc_Nzconst} the potential energy curves for three of those systems.
The $^{178}$Pt is the system for which the gap between the symmetric and asymmetric energy around scission ($Q_{20} \simeq$~90 b) is the most important (about 5 MeV). It benefits from the shell-gaps at Z~=~34 and N~=~56 that reduce the energy of the asymmetric mode.
For the $^{170}$W, this difference is only $\sim$2.5 MeV, which could be due to the fact that the $N_H$~$\simeq$~54 and $Z_L$~$\simeq$~32.5 values deviate slightly from 56 and 34. Nevertheless, asymmetric fission is still predicted to dominate in this system.
The case of $^{196}$Rn is interesting because the symmetric path becomes more energetically favorable than the asymmetric one. The asymmetric mode has $N_H \simeq$ ~60 and $Z_L \simeq$~38.5 values that deviate significantly from 56 and 34. The symmetric path, however, produces fragments with N~$\simeq$~55 which is close to N~=~56. Compact octupole deformed shell effects are then expected to favour symmetric fission in this nucleus.

\begin{figure}[h]
\includegraphics[width= 1. \linewidth]{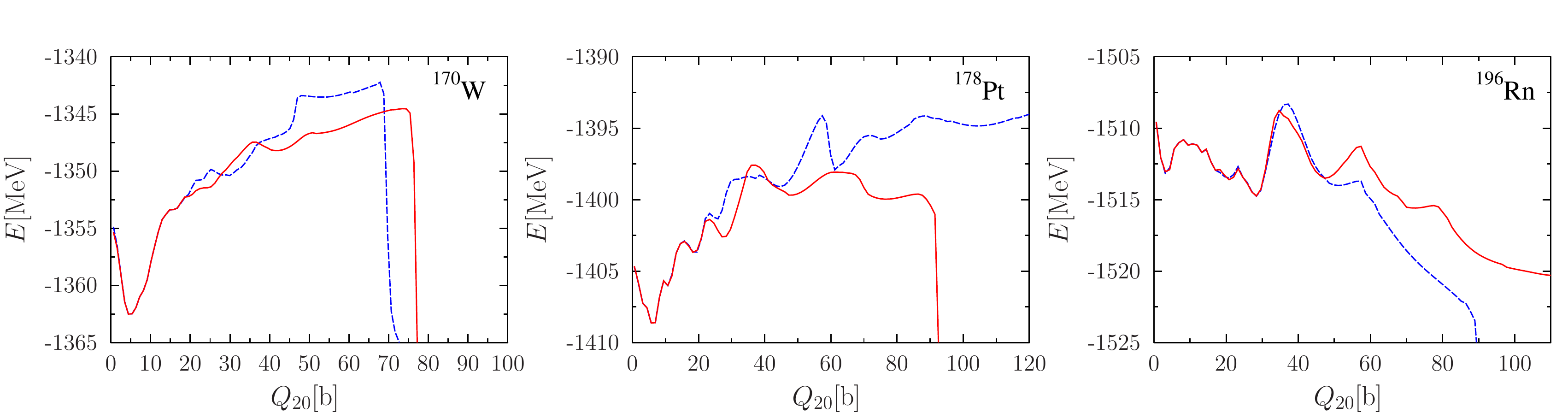}
\caption{Potential energy as a function of quadrupole moment
along the symmetric fission path (dashed line) and asymmetric fission valley (solid line) of $^{170}$W, $^{178}$Pt, and $^{196}$Rn.}
\label{fig:PES_proc_Nzconst}
\end{figure}

\section*{Summary}
 
CHF+BCS calculations have been performed to determine the main fission asymmety valley in the $^{180}$Hg region. 
We conclude  that, with the sly4d functional and our choice of pairing interaction, the N~=~56 is stronger to fix the asymmetry than the Z~=~34. 

 \end{widetext}

\end{document}